\documentclass[journal,transmag]{IEEEtran}

\usepackage{amssymb}
\usepackage{amsmath}
\usepackage{array}
\usepackage{color}
\usepackage{xcolor}
\usepackage{graphicx}
\usepackage{epstopdf}
\usepackage{epigraph}
\usepackage{hyperref}
\hypersetup{
	colorlinks=true,
	citecolor = blue,
	linkcolor=blue,
	filecolor=magenta,      
	urlcolor=cyan,
}
\usepackage{cite}

\newtheorem{example}{Example}

\newcommand\figref[1]{Fig.~\ref{#1}}
\newcommand\tableref[1]{Table~\ref{#1}}

\newcommand\sectref[1]{Section~\ref{#1}}

\newcommand{\iu}   {i}     
\newcommand{\Deltarm}   {\mathrm{\Delta}}

\newcommand{\Omegarm}   {\mathrm{\Omega}}

\newcommand{\Psirm}     {\mathrm{\Psi}}

\newcommand{\bfH}   {\mathbf{H}}
\newcommand{\bfE}   {\mathbf{E}}
\newcommand{\bfD}   {\mathbf{D}}

\newcommand{\bfr}   {\mathbf{r}}

\newcommand{\calN}  {\mathcal{N}}
\newcommand{\calT}  {\mathcal{T}}

\begin{document}
	\title{Accuracy of Difference Schemes in Electromagnetic Applications:\\
		a Trefftz Analysis}
	\author{\IEEEauthorblockN{Igor Tsukerman\IEEEauthorrefmark{1}
		}
		\IEEEauthorblockA{\IEEEauthorrefmark{1}Department of Electrical and 
		Computer Engineering,\\
			The University of Akron, OH 44325-3904, USA,
			igor@uakron.edu\thanks{Research was supported
				in part by the US National Science Foundation awards 
				DMS-1216927 and DMS-1620112.}}
	}
	
\IEEEtitleabstractindextext{%
\begin{abstract}
	The paper examines \textit{local} approximation errors
	of finite difference schemes in electromagnetic analysis.
	Despite a long history of the subject, several 
	accuracy-related issues have been overlooked and/or remain 
	controversial. For example, conflicting claims have been made
	in the literature about the order of Yee-like schemes 
	in the vicinity	of slanted or curved material interfaces.
	Two novel practical methods for comparison of local accuracy
	of difference schemes are proposed: one makes use of 
	\textit{Trefftz test matrices}, and the other one
	relies on a new measure of approximation accuracy in 
	\textit{scheme-exact Trefftz subspaces}.
	One particular conclusion is that a loss of accuracy 
	for Yee-like schemes at slanted material
	boundaries is unavoidable.
\end{abstract}
		
\begin{IEEEkeywords}
		Difference schemes, finite difference time domain,
		approximation accuracy, material interfaces, Trefftz functions,
		Trefftz spaces.
\end{IEEEkeywords}}
	
\maketitle
\thispagestyle{empty}
\pagestyle{empty} 
	
\section{Introduction}\label{sec:Intro}
%
Finite difference (FD) schemes have been known for over a century
\cite{Thomee01} and in computational electrodymanics
have been extensively used since the publication of Yee's paper \cite{Yee66}
in 1966.
Despite this long history, several principal issues 
related to the accuracy of various schemes
are not yet resolved to the satisfaction of application
engineers, developers and practitioners. 
These issues appear in the bulleted list of 
\sectref{sec:Factors-affecting-analysis};
but to motivate the discussion, let us start with the
following observations.

The Yee schemes on regular Cartesian grids are well known to be
of second order in a homogeneous medium
and also at material interfaces \textit{running along the grid lines}
\cite{Hirono-Shibata00,Hwang-Cangellaris01}.
In the vicinity of oblique or curved interfaces, however,
the order of difference schemes 
is not as easy to evaluate. 
The problem is exacerbated by the availability
of various interpolation schemes for staggered grids 
\cite{Shyroki11} and the difference in the numerical
accuracy of \textit{local} and \textit{global} quantities. In the engineering 
literature,
analysis is rarely performed with full mathematical rigor, so
ambiguities do arise. A typical and important example is the ``subpixel
smoothing'' technique, which in \cite{Farjadpour06,Oskooi09}
is claimed to be of second order,
but according to \cite{Bauer-Werner-Cary-FDTD11}
``... has first-order error (possibly obscured up to high resolutions or high
dielectric contrasts)''. Claims about second-order accuracy have also been made 
with regard to Finite Integration Techniques
(FIT) \cite{Clemens02}.

Numerical evidence for second-order convergence of Yee-like schemes 
with a particular type of interpolation
between different $\bfE$ and $\bfD$ field components on staggered grids
is provided in \cite{Shyroki11}. 
However, this convergence is with respect to a \textit{global} quantity 
(resonance frequency) rather than the local fields near a slanted 
or curved interface.
%
\section{Factors Affecting the Accuracy 
Analysis}\label{sec:Factors-affecting-analysis}
%
The inconsistencies noted above can be attributed to several factors:
\begin{itemize}
	\item difference in the numerical
	accuracy of local and global quantities;
	\item availability of various interpolation schemes 
	for staggered grids \cite{Shyroki11};
	\item scaling: multiplication of any scheme by an arbitrary factor 
	(possibly dependent on the grid size and/or time step)
	changes the consistency error.	
\end{itemize}
The last item can be illustrated with a very simple example.
For the 2D Laplace equation, the
standard five-point stencil (set of FD coefficients)
on a uniform Cartesian grid
is $\{-1, -1, 4, -1, -1 \} \, h^{-2}$, 
where 4 corresponds to the central node; this scheme is well known
to be of second order. At the same time,
the finite element -- Galerkin procedure on a regular mesh 
with linear triangular elements produces the same scheme 
but without the $h^{-2}$ factor, which formally makes it 
a scheme of order 4 (!?).

Needless to say, rescaling of the scheme does not affect the FD solution
(in the absence of round-off errors); rather, it alters the balance
between the consistency and stability estimates in the
Lax-Richtmyer theorem, as outlined later in this section.
Thus a formal way to fix the ``true'' order of a scheme would be
to restrict consideration to schemes for which the stability constant 
is $\mathcal{O}(1)$. In practice, though, one typically wishes
to separate the issues of consistency and stability to the extent possible,
because the former lends itself to analysis much more easily
than the latter. Moreover, \textit{local} error estimates
are of great practical interest, whereas the Lax-Richtmyer theorem
relates only global quantities.

Many standard schemes feature a 1:1 correspondence between derivatives
in the original differential equation and the respective FD terms
(e.g. $d_x^2 \leftrightarrow \{1, -2, 1\} h^{-2}$),
which makes the proper scaling intuitively clear.
However, this direct correspondence may not be quite as obvious
or may not even exist in other cases -- notably, for FLAME
schemes \cite{Tsukerman05,Tsukerman06,Tsukerman-PBG08,
	Cajko-NIM-FLAME08,Tsukerman-JCP10}.
To illustrate this point, consider as an example the high-order
$3 \times 3$ FD stencil for the 2D Helmholtz equation
$\nabla^2 u + k^2 u = 0$ with, for simplicity, a real positive wavenumber $k$
(\cite[p.~2218]{Tsukerman05}, \cite[p.~695]{Tsukerman06},
 \cite[p.~1379]{Cajko-NIM-FLAME08}, 
 and a similar scheme of \cite[p.~342]{Babuska-Ihlenburg95}):
  
\begin{figure}
	\centering
	\includegraphics[width=0.5\linewidth]{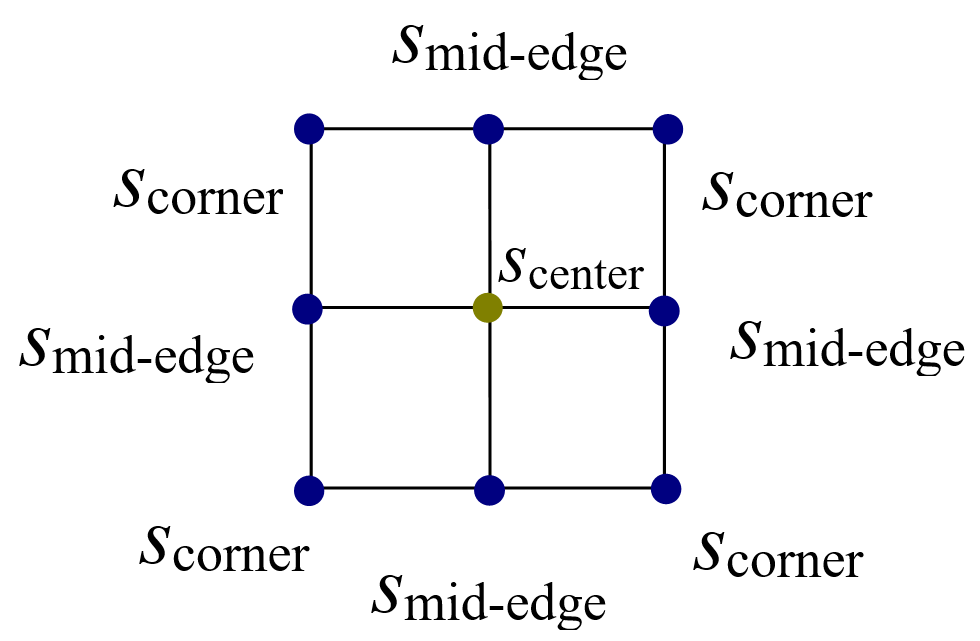}
	\caption{A schematic representation of the 
		FLAME scheme \eqref{eqn:FLAME_coeffs}
		for the 2D Helmholtz equation in free space. Grid size $h$
		in both coordinate directions.}
	\label{fig:FLAME-scheme-9pt-stencil}
\end{figure}

\begin{equation}\label{eqn:FLAME_coeffs}
    s_{\mathrm{center}} = \frac{A}{D} \left( e_{\frac12}+1 \right), 
    ~~ s_{\mathrm{mid-edge}} = -\frac{B}{D}, ~~ 
    s_{\mathrm{corner}} = \frac{C}{D} \, e_{-\frac12}
\end{equation}
where
$$
A ~=~ e_{\frac{1}{2}} e_1  +  2 e_{\frac{1}{2}} e_0  -  4 e_{-\frac{1}{2}} 
e_1 + e_{\frac{1}{2}} -  4 e_{-\frac{1}{2}}  +  e_1  +  2e_0  +  1  
$$
$$
B ~=~ e_{\frac{3}{2}} e_0  -  2 e_{\frac{1}{2}} e_1  +  2 e_{\frac{1}{2}} 
e_0-2e_{\frac{1}{2}}  +  e_0
$$
$$
C ~=~ 2 e_{\frac{1}{2}} e_0  -  e_{-\frac{1}{2}} e_1  -  2 e_{-\frac{1}{2}} 
e_0  -  e_{-\frac{1}{2}}  +  2 e_0
$$
$$
D ~=~ (e_0-1)^2(e_{-\frac12}-1)^4;
$$
$$
 e_\gamma \,=\, \exp(2^\gamma ihk), \quad
 \gamma \,=\, -\frac12, 0, \frac12, 1, \frac{3}{2}
$$
This stencil is schematically represented in
 \figref{fig:FLAME-scheme-9pt-stencil}.

For \eqref{eqn:FLAME_coeffs} and similar schemes, it is not immediately
obvious what the ``right'' scaling should be (whether multiplication
by a certain power of $h$ would be appropriate).
In a comprehensive convergence theory, which would include stability analysis
in addition to approximation errors, the scaling issue would 
not be critical for the analysis of \textit{global} errors, 
but would still matter for the assessment of \textit{local} ones. 
In any event, since stability is much more difficult to study than 
approximation,
it is desirable to establish a universal measure by which the
local approximation accuracy of difference schemes can be compared
in a consistent manner.

Godunov \& Ryabenkii (G\&R) discuss a closely related 
but different issue \cite[Section~5.13]{Godunov87}. 
They note that consistency error can be formally reduced by changing the norm 
in which this error is measured. Incorporating an arbitrary power of $h$
into that norm\footnote{For simplicity
	of notation, we consider only one parameter $h$ here; the statement can
	obviously be generalized to several grid sizes in different 
	coordinate directions and in time.} 
or, even more dramatically, a factor like $2^{-1/h}$, one could
magically ``improve'' the accuracy of the scheme, although the numerical
solution would remain unchanged. Given a discrete space $U_h$ where
an FD scheme ``lives'' and the corresponding continuous space
$U$ for the underlying boundary value problem, G\&R write:
\begin{quotation}
	It is customary to choose a norm in the space $U_h$ 
	in such a way that,	as $h$ tends to zero, it will go over into 
	some norm for functions given on the whole [domain], i.e. so 
	that
	\begin{equation}\label{eqn:Uh-norm-tends-to-U-norm}
	   \lim_{h \rightarrow 0} \| u_h \|_{U_h} \,=\,
	    \| u \|_U
	\end{equation}
\end{quotation}
Trivial examples of that are (i) the maximum norms in both spaces;
(ii) the discrete norm $\| u_h \|_{U_h}^2 = \left( h^d \sum_m |u_m|^2 \right)$,
which transforms into the $L_2$ norm as $h \rightarrow 0$.

G\&R's condition \eqref{eqn:Uh-norm-tends-to-U-norm} is natural
but \textit{separate from the scaling issue}. 
Indeed, even if the norms in both
discrete and continuous spaces are fixed (say, for the sake of argument,
taken to be the maximum-norms), the scheme can still be rescaled 
by an arbitrary factor, including an $h$-dependent factor.

To proceed further, we need to recall the Lax-Richtmyer theorem 
relating consistency, stability 
and convergence \cite{Strikwerda04,Godunov87}.
The connection is easy to see if the difference systems 
for the numerical
and exact solutions are written side by side:
$$
S_h \underline{u}_h \,=\, \underline{f}_h;  
\quad
S_h \underline{u}^*_h \,=\, \underline{f}_h + \underline{\epsilon}_c
$$
Subtracting these equations, one immediately observes that
\begin{equation}\label{eqn:S-uh-minus-ustar-eq-epsc}
S_h \underline{\epsilon}_s \,=\, \underline{\epsilon}_c,  ~~~
\underline{\epsilon}_s \equiv \underline{u}_h - \underline{u}^*_h
\end{equation}
or equivalently
\begin{equation}\label{eqn:epss-eq-inv-S-epsc}
     \underline{\epsilon}_s \,=\, S_h^{-1}  \underline{\epsilon}_c
\end{equation}
Here $\underline{\epsilon}_s \in \mathbb{C}^n$ (or $\mathbb{R}^n$,
depending on the type of the problem) is the solution 
error \textit{vector}. This is a simple but critical result relating 
consistency and solution errors. It is important to note,
however, that all the quantities in \eqref{eqn:epss-eq-inv-S-epsc}
are \textit{global}.
%
\section{Accuracy of Yee-like Schemes:\\
	General Considerations}\label{sec:Accuracy-Yee-general}
%
We can now take a closer look 
at the accuracy of Yee-like schemes\footnote{The term ``Yee-like'' includes, 
	for brevity, FIT \cite{Clemens02} but excludes from immediate consideration 
	FEM-FD hybrids, complex meshes, subcell divisions, etc.
	\cite{Schauer-partial-cells03,DeGersem-FIT-FE08,Zagorodnov03}.
	Still, the methodology of the paper can be applied to all schemes.}
near slanted material boundaries. Consider an FD scheme in the 
following generic form, in the absence of sources:
\begin{equation}\label{eqn:u-next-eq-Su}
   \underline{u}^{(i)T} \underline{s}^{(i)} 
   (h, \Deltarm t, \epsilon, \mu)  \,=\, 0
\end{equation}
where $\underline{u}^{(i)}$ is a Euclidean vector of degrees of freedom
corresponding to a particular grid ``molecule''\footnote{This 
	locution was apparently coined by J.~P.~Webb
	\cite{Pinheiro09}. The set of \textit{coefficients}
	of a scheme is commonly referred to as a ``stencil'';
	surprisingly, however, there does not seem to be
	a standard term for the \textit{set of nodes} over which an FD scheme
	is defined.}  $i$ with node locations
$\bfr_{\beta}^{(i)}, t_{\beta}^{(i)}$ ($\beta = 1,2,\ldots, n$), 
and $\underline{s}$ 
is the coefficient vector defining the scheme. As indicated in
\eqref{eqn:u-next-eq-Su}, these coefficients depend on the electromagnetic
parameters $\epsilon = \epsilon(\bfr)$ and $\mu = \mu(\bfr)$
(assuming linear characteristics of all media).
It is convenient to introduce a small local domain $\Omegarm^{(i)}$
containing the grid molecule (e.g. the convex hull
of the nodes of the molecule); diam~$\Omegarm^{(i)}$ is thus of the order
of the grid size.

The \textit{local} (stencil-wise) consistency error 
$\epsilon_c^{(i)}$ is defined as
\begin{equation}\label{eqn:epsc-eq-u-s}
    \epsilon_c^{(i)}(h, \Deltarm t, \epsilon, \mu)
    \,=\,
    \underline{s}^{(i)T} \calN^{(i)} u^* 
\end{equation}
In this expression,
$\underline{u}^{*} $ could in principle be any vector defined
on the grid molecule,
but is almost always taken to be the column vector of nodal values 
of the exact solution $u^*$ of the underlying boundary value problem.
Formally, therefore, $u^* \in \calT(\Omegarm^{(i)})$,
where $\calT(\Omegarm^{(i)})$ is the local \textit{Trefftz space} --
that is, the space of functions satisfying the underlying weak-form
differential equations in $\Omegarm^{(i)}$.
Finally, the operator 
$\calN^{(i)}: \calT(\Omegarm^{(i)}) \rightarrow \mathbb{R}^n$
(or $\mathbb{C}^n$) in \eqref{eqn:epss-eq-inv-S-epsc}
produces $n$ given degrees of freedom (typically, the nodal values)
of any given local solution $u^*$.

Our goal is to establish an accuracy measure which, unlike
$\epsilon_c^{(i)}$ \eqref{eqn:epsc-eq-u-s}, would be independent
of the scaling of the scheme $\underline{s}^{(i)}$.
Obviously, the numerical solution does not,
in the absence of roundoff errors, depend on this scaling.
(If all stencils were to be simultaneously rescaled by any nonzero factor
$\kappa$, the inverse matrix $S_h^{-1}$ in \eqref{eqn:epss-eq-inv-S-epsc}
would also be simultaneously rescaled, but by the inverse factor $\kappa^{-1}$.)
Although this is in principle clear, theoretical analysis 
faces several impediments -- partly mathematical and partly practical:
\begin{enumerate}
	\item Rigorous analytical estimates of the stability factor 
	related to the norm of $S^{-1}$ are available for fairly simple
	model problems but may be difficult or impossible to obtain 
	for more complicated practical ones. It is therefore desirable 
	to find a meaningful way to characterize approximation independently
	of stability.
	\item The Lax-Richtmyer estimate \eqref{eqn:S-uh-minus-ustar-eq-epsc}
	is global and does not connect the \textit{local} 
	solution errors to the respective \textit{local} consistency errors.
	Such local estimates are available for finite element methods,
	thanks to variational formulations and duality principles
	\cite{Schatz-Wahlbin77,Schatz-Wahlbin78,Schatz-Wahlbin79,Schatz-Wahlbin95,
		Schatz-Thomee-Wahlbin98,Demlow-localized-estimates07,
		Demlow-FEM-approximation12}, but not for FD schemes.
		\label{list:FD-scaling-problem}
	\item To reiterate, stability is, by its nature, a \textit{global} feature;
	that is, it pertains to the system of FD equations as a whole.
	At the same time, \textit{local} approximation errors are of interest
	in their own right. In particular, one may wish to compare
	the local accuracy of two schemes, possibly
	derived from completely different considerations
	and involving disparate degrees of freedom.
\end{enumerate}
Two general and scaling-independent ways of examining 
the \textit{local} errors are introduced in 
Sections~\ref{sec:Trefftz-Test-Matrix} and 
\ref{sec:Approximation-Trefftz-subspaces} below.
%
\section{Trefftz Test Matrix and Its\\
	Minimum Singular Value}
\label{sec:Trefftz-Test-Matrix}
%
Obviously, for any single solution $u^*$ one can generate infinitely many
exact schemes -- that is, schemes with a zero consistency error
$\epsilon_c$ \eqref{eqn:epsc-eq-u-s}. 
Since only one FD stencil is considered in the remainder of this section,
the superscript $(i)$ indicating the stencil number is now dropped.
Let the position vectors of $n$ stencil nodes be $\bfr_{\beta}$,
$\beta = 1,2,,\ldots, n$.

Suppose that we have a Trefftz test set of $m$ different 
solutions $u^*_{1,\ldots,m}(\bfr)$ independent of the mesh size and time step.
By definition, \textit{Trefftz functions} -- 
in particular, $u^*_{\alpha}(\bfr)$ -- satisfy
the differential equation
of the problem and interface boundary conditions (if any) within the convex hull
of the grid molecule.
We want the consistency error to be small \textit{for each test solution}. 

Let us arrange the nodal values\footnote{If degrees of freedom
	(DoF) other than nodal values are employed in the FD scheme,
	then these DoF will appear in the right hand side of
	\eqref{eqn:U-alpha-beta-eq-u} in the place of $u^*_{\alpha} 
	(\bfr_{\beta})$.} 
of $u^*_{1,\ldots,m}(\bfr)$
in a matrix $U$: each row $\alpha$ of $U$ contains the nodal values of 
the test solution $u^*_{\alpha}(\bfr)$, so that
\begin{equation}\label{eqn:U-alpha-beta-eq-u} 
     U_{\alpha \beta} ~=~ u^*_{\alpha} (\bfr_{\beta})
\end{equation}
Assembling the consistency errors \eqref{eqn:epsc-eq-u-s} for all
test solutions into one Euclidean vector 
$\underline{\epsilon}_c \in \mathbb{R}^m$, we get
\begin{equation}\label{eqn:epsc-geq-sigma-s}
    \underline{\epsilon}_c \, \equiv \,  U \underline{s}
    \quad ~~ \Rightarrow \quad ~~ 
    \| \underline{\epsilon}_c \|_2 \, \geq \,
    \sigma_{\min} (U) \| \underline{s} \|_2
\end{equation}
where $\sigma_{\min}(U)$ is the minimum singular value of 
$U = U(h, \Deltarm t, \epsilon, \mu)$.
This is a lower bound of the consistency error. 
Using different sets of Trefftz functions,
or one large set, one can obtain different bounds of this form.

A rigorous mathematical solution of the stability and scaling 
problems is not
in general available, but the error bound \eqref{eqn:epsc-geq-sigma-s}
suggests an alternative on physical/engineering grounds.
Namely, one can compare the minimum singular values of matrix
$U(h, \Deltarm t, \epsilon, \mu)$ and its free-space instantiation
$U_1(h, \Deltarm t) = U(h, \Deltarm t, \epsilon = 1, \mu = 1)$:
\begin{equation}\label{eqn:sigma-U-sigma-U1}
    \zeta \,=\, \frac{ \sigma_{\min} (U(h, \Deltarm t, \epsilon, \mu))}
    {\sigma_{\min} (U(h, \Deltarm t, 1, 1))}
\end{equation}
Clearly, in the inhomogeneous case one may expect 
the estimate \eqref{eqn:epsc-geq-sigma-s} to be worse than
the respective one for free space; hence typically $\zeta > 1$
and quite possibly $\zeta \gg 1$.
Let us consider illustrative examples.

\begin{example} 
	To start with, we calculate just the free-space value 
	$\sigma_{\min} (U_1)$ for the 2D Laplace equation $\nabla^2 u = 0$. 
	Introduce the standard five-point stencil and assume, 
	for simplicity of notation,
the same grid size $h$ in both coordinate directions; the node coordinates are
$$
x_{1,\ldots,5} = \{0, 0, -h, h, 0\};~~~
y_{1,\ldots,5} = \{0, -h, 0, 0, h\}
$$
(thus, node \#1 is the central node in this grid molecule).
Let the Trefftz basis $\{ u^*_\alpha(x,y) \}$
consist of harmonic polynomials up to order 4
(these are the real and imaginary parts of $(x + \iu y)^k, k = 0,\ldots, 4$):
{\small
$$
u^*_{1,\ldots,8} = 
\{ 1, x, y, x^2 - y^2, 2xy, x^3 - 3xy^2, 3x^2y - y^3, x^4 - 6x^2 y^2 + y^4 \}
$$
}
Then the $U$ matrix is
{\small
$$
U = \begin{pmatrix}
1  ~&~ 0  ~&~ 0  ~&~ 0  ~&~ 0  ~&~ 0  ~&~ 0  ~&~ 0\\ 
1  ~&~ 0  ~&~ -h  ~&~ -h^2  ~&~ 0  ~&~ 0  ~&~ h^3  ~&~ h^4\\ 
1  ~&~ -h  ~&~ 0  ~&~ h^2  ~&~ 0  ~&~ -h^3  ~&~ 0  ~&~ h^4\\ 
1  ~&~ h  ~&~ 0  ~&~ h^2  ~&~ 0  ~&~ h^3  ~&~ 0  ~&~ h^4\\ 
1  ~&~ 0  ~&~ h  ~&~ -h^2  ~&~ 0  ~&~ 0  ~&~ -h^3  ~&~ h^4 
\end{pmatrix}
$$
}
\noindent
(Recall that each column of this matrix contains the nodal values of the
respective test function. Obviously, function $2xy$, corresponding
to the zero column, could have been omitted from the basis.)

Symbolic algebra gives the following lowest order terms for the singular
values of $U$:
$$
\sigma_{1,\ldots,5}(N) \sim \{ 2 h^2, \, 2^{\frac12} h, \, 2^{\frac12} h,
\, 2 \cdot 5^{-\frac12} h^4, \, 5^{\frac12} \}
$$
Hence
$$
\sigma_{\min}(U) \overset{h \ll 1}{=} 
\sigma_{4}(U) \sim 2 \cdot 5^{-\frac12} h^4
= \mathcal{O}(h^4)
$$
\end{example}

\begin{example}
Consider now the 2D wave equation 
$$
v^2 \nabla^2 u - \partial_t^2 u = 0
$$ 
We follow the same exact steps as in the previous example.

The simplest grid ``molecule'' has seven nodes in the $xyt$ space:
$$
x_{1,\ldots,7} = \{0, 0, -h, h, 0, 0, 0\}
$$
$$
y_{1,\ldots,7} = \{0, -h, 0, 0, h, 0, 0\}
$$
$$
t_{1,\ldots,7} = \{0, 0, 0, 0, 0, -\Deltarm t, \Deltarm t\};
$$
where the Trefftz test set $\{ u_\alpha(x,y) \}$ is now chosen as
\begin{equation}\label{eqn:u-eq-x-pm-vt-k} 
    u^*_{1,\ldots,17} = \{1,  (x \pm vt)^k,  (y \pm vt)^k \}, ~~~ k = 1,2,3,4
\end{equation}
For the $U$ matrix corresponding to the chosen grid molecule and test set
 \eqref{eqn:u-eq-x-pm-vt-k}, symbolic algebra yields
\begin{equation}\label{eqn:sigma-min-h4}
   \sigma_{\min}(U) \overset{h \ll 1}{\sim} c_4 h^4
\end{equation}
The analytical expression for the coefficient $c_4$ is too cumbersome to be
reproduced here, but it is plotted, as a function of $v \Deltarm t / h$, 
in \figref{fig:c4-sigma-min-vs-vdth}.
\begin{figure}
	\centering
	\includegraphics[width=0.75\linewidth]{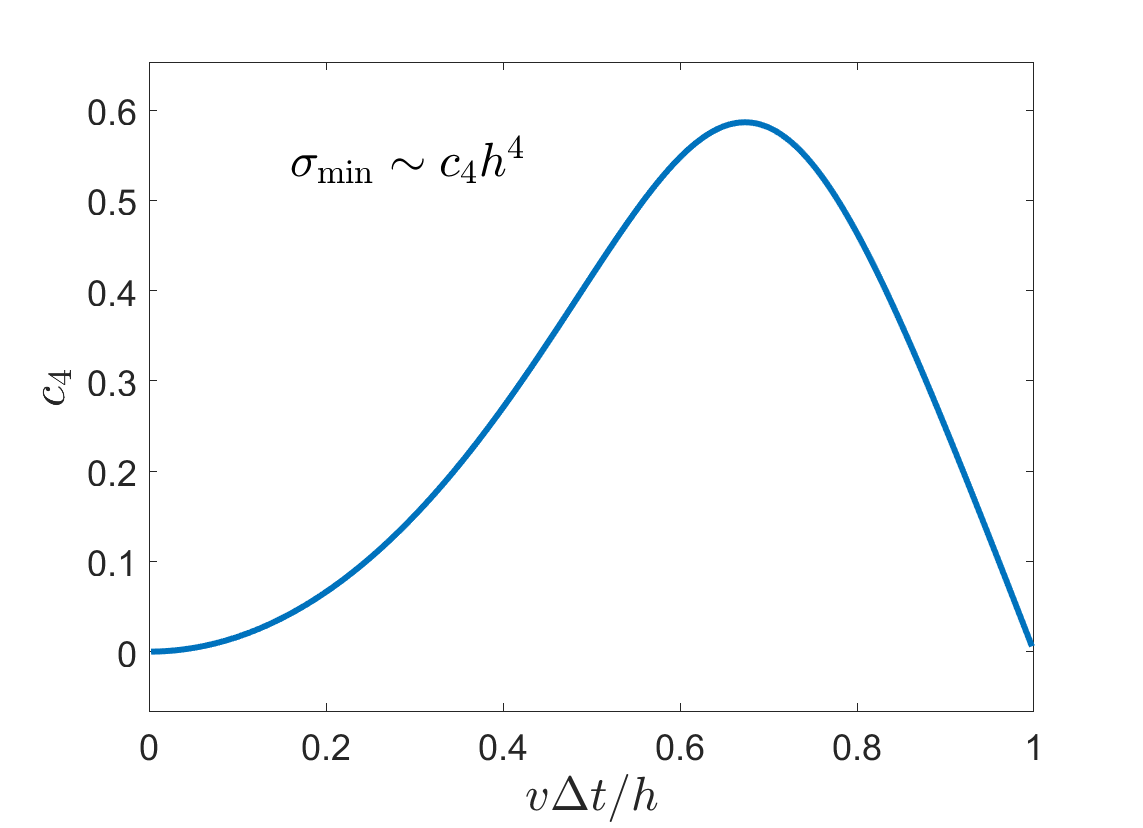}
	\caption{Coefficient $c_4$ of \eqref{eqn:sigma-min-h4}
		vs. $v \Deltarm t / h$, where 
		$\sigma_{\min}(U) \sim c_4 h^4$.}
	\label{fig:c4-sigma-min-vs-vdth}
\end{figure}
\end{example}

\begin{example}
\label{page:FDTD-order-main-example}
Following the preliminary examples above, we are in a position
to examine the Yee setup in 2D in a similar fashion. 
Consider two dielectric media with permittivities
$\epsilon_{1,2}$, separated by a slanted interface boundary 
(\figref{fig:EH-grid-FDTD=2D-interface}). Of most interest is the $H$-mode
($p$-mode), with a one-component $\bfH$ field perpendicular to the plane 
of the figure and a two-component $\bfE$ field whose normal component
is discontinuous across the interface. The $H$ nodes are labeled with 
the yellow spheres in the figure, and the $E_{x,y}$ components -- with the 
arrows. The time axis is for simplicity not shown, but it is understood
that the central $H$ sphere indicates in fact a triple Yee node 
($t_0, t_0 \pm \Deltarm t$), and that the $E_{x,y}$ arrows indicate 
double nodes ($t_0 \pm \Deltarm t / 2$); $t_0$ is any ``current'' 
moment of time. There are 7 degrees of freedom for the $H$ field
(five at $t_0$ and two at $t_0 \pm \Deltarm/2$) and $2 \times 4 = 8$
for $E_{x,y}$ -- altogether 15 degrees of freedom.

One horizontal and one vertical black lines in
\figref{fig:EH-grid-FDTD=2D-interface}
belong to the $H$ grid with a size $h$ in each direction;
the finer blue grid
indicates $h/4$ subdivisions as a visual aid. 
The slope $\theta$ of the interface is an adjustable parameter. 
For definiteness, we take $\theta = 30^{\circ}$ and assume that 
the interface boundary divides the ``horizontal'' segment between 
two adjacent $H$ nodes in the ratio of 1:3, as indicated in the figure.
With this geometric setup, two $H$ nodes (the left one and the top one)
happen to lie in the $\epsilon_1$ dielectric, as does the top ``double node''
$E_x$. All other nodes lie in the second dielectric.

\begin{figure}
	\centering
	\includegraphics[width=0.6\linewidth]{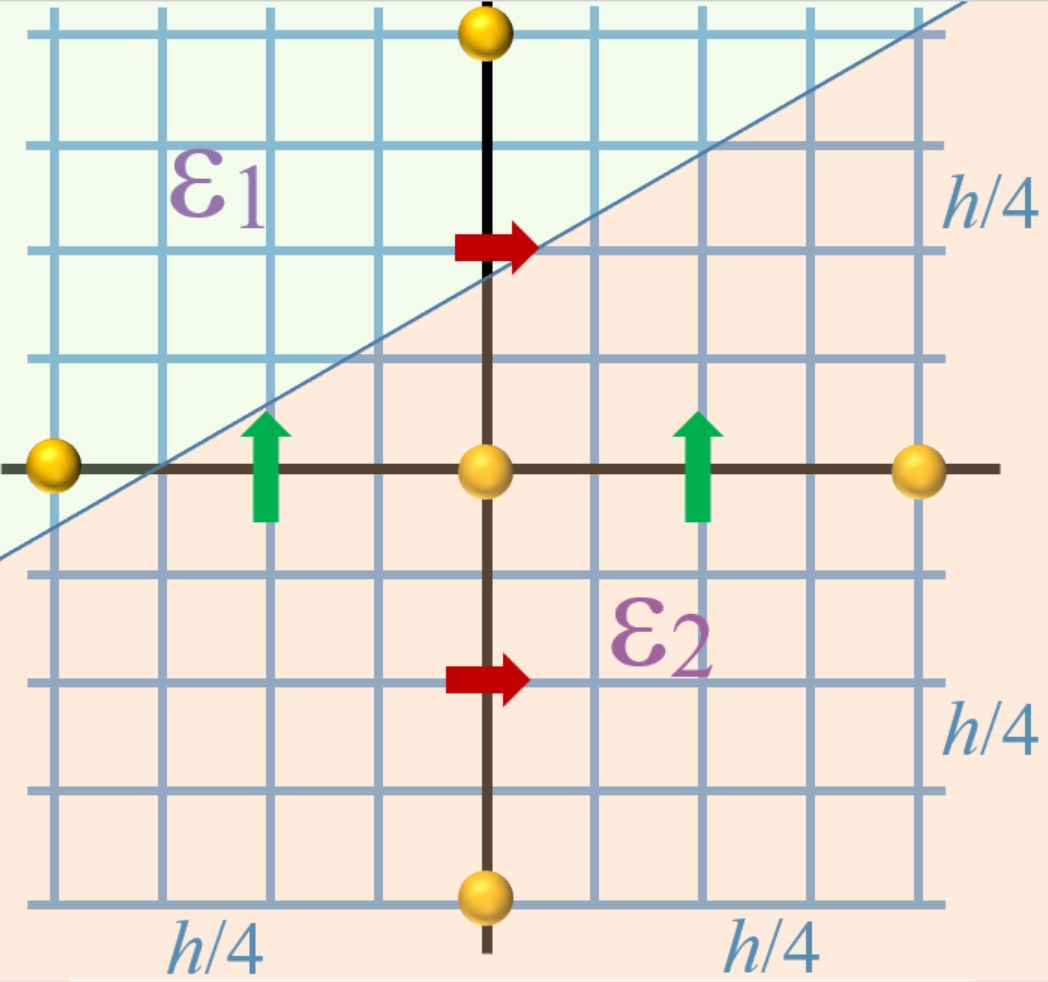}
	\caption{Yee-like schemes near material boundaries. $H$-mode ($p$-mode):
		a one-component $\bfH$ field perpendicular to the plane of the figure 
		and	
		a two-component $\bfE$ field in the plane. Two dielectric
		media ($\epsilon_{1,2}$) with an interface boundary slanted with respect
		to the Yee grid. Yellow spheres: $H$ nodes; horizontal arrows: $E_x$ 
		nodes; 
		vertical arrows: $E_y$ nodes. The central $H$ sphere indicates a triple
		node with respect to time ($t_0, t_0 \pm \Deltarm t$), and the 
		$E_{x,y}$ 
		arrows indicate double nodes ($t_0 \pm \Deltarm t / 2$); $t_0$ 
		is any ``current'' moment of time.}
	\label{fig:EH-grid-FDTD=2D-interface}
\end{figure}
The Trefftz test set is a slightly generalized version of
\eqref{eqn:u-eq-x-pm-vt-k}. Now that two media with their respective phase
velocities $v_{1,2}$ are present, each polynomial of the form 
$(x \pm v_1 t)^k$,  $(y \pm v_1 t)^k$ in the first subdomain is matched,
via the standard interface boundary conditions, with the corresponding 
polynomial $(x \pm v_2 t)^k$, $(y \pm v_2 t)^k$ in the second subdomain;
we take $k = 0,1,2,3$.

\begin{figure}
	\centering
	\includegraphics[width=0.95\linewidth]
	{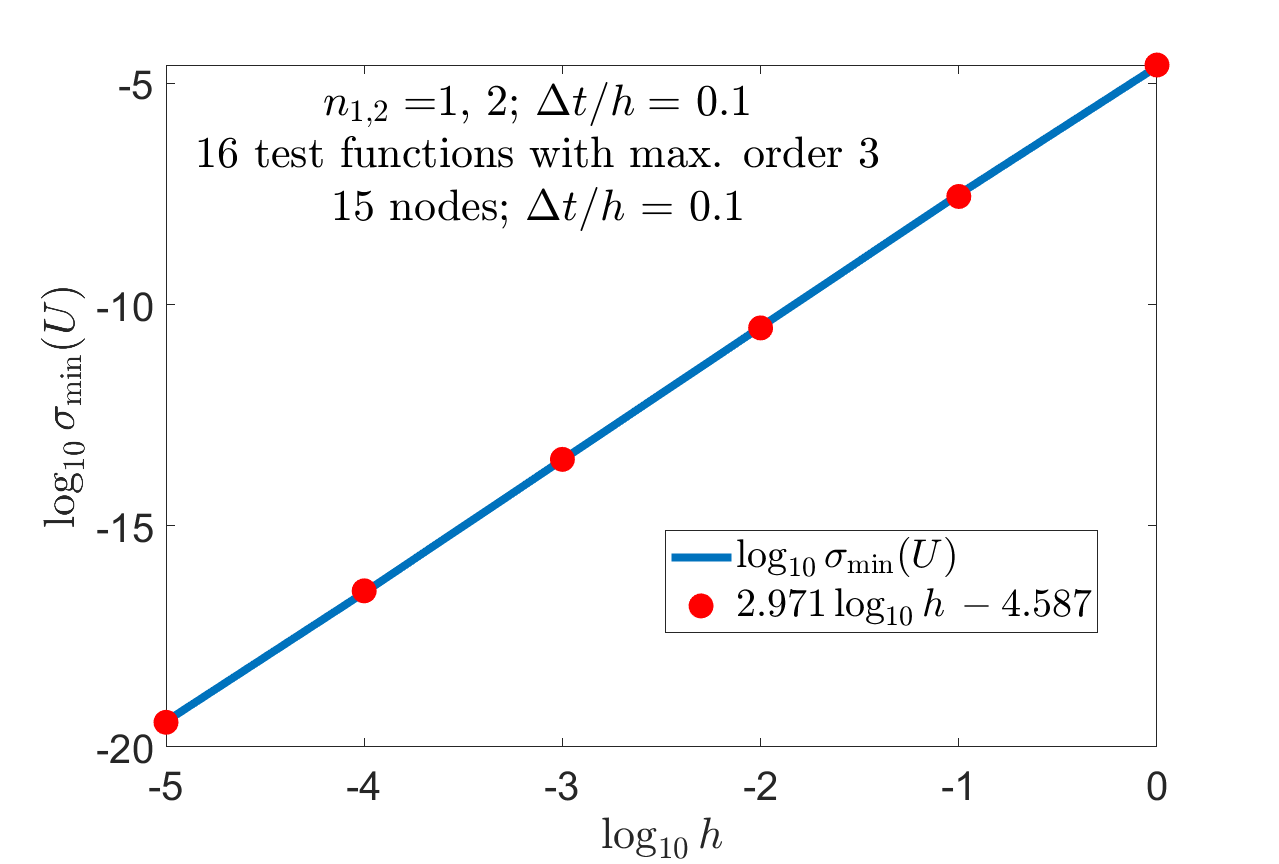}
	\caption{$\log_{10} \sigma_{\min}(N)$ vs. $\log_{10} h$.
		Solid line: $\log_{10} \sigma_{\min}(U)$; markers: linear fit with
		slope K$\approx 3$. Indexes of refraction
		$n_{1,2} =$1, 2.  $c \mathrm{\Delta} t / h = $ 0.1.
		Geometric setup of p.~\pageref{page:FDTD-order-main-example};
		material interface at angle $\theta = \pi/$6.
		15-node stencil: 7 nodes for $H$, $2 \times 2 = 4$ nodes for $E_x$, 
		$2 \times 2 = 4$  nodes for $E_y$. Trefftz test basis:
		polynomial traveling waves in each dielectric, matched via
		the standard interface boundary conditions (see text).}
	\label{fig:sigma-min-vs-h-EH-stencil-interface-16tf-order3}
\end{figure}

The rest of the analysis proceeds the same way as in the previous
examples, except that a final closed-form analytical result for
$\sigma_{\min}(U)$, where $U$ is in this case a $15 \times 16$ matrix,
is no longer feasible to obtain via symbolic algebra. Instead,
this smallest singular value is computed in variable precision arithmetic
(32 digits). The results are plotted in 
\figref{fig:sigma-min-vs-h-EH-stencil-interface-16tf-order3}
and clearly indicate that
\begin{equation}\label{eqn:FDTD-sigma-min-U-Yee-interface}
     \sigma_{\min}(U) \,=\, \mathcal{O}(h^3)
     \quad \left( \mathrm{fixed} ~ \frac{c \Deltarm t}{h} \right)
\end{equation}
At the same time, in free space
\begin{equation}\label{eqn:FDTD-sigma-min-U-Yee}
   \sigma_{\min}(U) \,=\, \mathcal{O}(h^4)
   \quad \left( \mathrm{fixed} ~ \frac{c \Deltarm t}{h} \right)
\end{equation}
That is, in the presence of a dielectric interface one cannot
achieve the same order of convergence of Yee-like schemes 
as in free space, and the deterioration factor $\zeta$
\eqref{eqn:sigma-U-sigma-U1} is
\begin{equation}\label{eqn:sigma-U-sigma-U1-Yee}
   \zeta_{\mathrm{Yee}} \,=\, \mathcal{O} \left( h^{-3} / h^{-4} \right)
   \,=\, \mathcal{O} (h)
\end{equation}
This conclusion is not entirely unexpected. Its limitation 
is that this analysis has been applied to specific degrees
of freedom adopted in classical Yee schemes -- that is, the nodal values of the
fields on staggered grids. A generalization
to other degrees of freedom -- notably, to edge circulations 
and/or surface fluxes --
and to non-staggered grids\footnote{Such grids are used e.g. in
	pseudospectral time domain methods (PSTD) \cite{Liu-PSTD97}.}
is certainly possible.
However, the odds of preserving the second order of the free-space Yee scheme
in the presence of interfaces are slim;
see also \cite{Tornberg-Engquist08}. 
On the positive side, as already noted, second-order schemes are available 
for interface boundaries parallel to the gridlines,
and there is also compelling evidence
that \textit{global} quantities can also be evaluated with second-order accuracy
if FDTD schemes are constructed judiciously 
\cite{Bauer-Werner-Cary-FDTD11,Shyroki11}.
\end{example}
%
\section{Approximation in Trefftz Subspaces}
\label{sec:Approximation-Trefftz-subspaces}
%
Recall that consistency error, by its standard definition
\eqref{eqn:epsc-eq-u-s}, depends on the scaling of the underlying scheme.
Namely, multiplication of the scheme by an arbitrary
factor, including a factor depending on the grid size $h$
and/or the time step $\Deltarm t$,
leads to a commensurate change of the consistency error.
A related observation is that FEM on regular grids
often produces essentially the same schemes as standard FD,
except for an extra factor of the order $\mathcal{O}(h^d)$ due to integration
($d = 1,2,3$ is the number of dimensions). Finally, schemes like
FIT, where degrees of freedom
are field circulations and fluxes rather than nodal values,
also involve inherently different scaling factors. 

Notably, in  finite element analysis the focus is on 
\textit{continuous-level} approximation rather than 
FD-type consistency errors.
Arguably, the former is more fundamental than the latter --
in particular, continuous-level errors do not depend on 
any scaling of the respective FE equations. More precisely,
given a finite element $K$, a suitable functional space $X = X(K)$,
a function $u \in X(K)$ and, importantly,
a finite-dimensional subspace $X_h = X_h(K) \subset X(K)$,
the approximation error for $u$ is unambiguously defined as 
\begin{equation}\label{eqn:eps-a-FEM}
   \epsilon_a \,\overset{\mathrm{def}}{=}
   \inf_{u_h \in X_h} \| u - u_h \|
\end{equation}
With this definition, the error is completely decoupled from
any algebraic manipulation of the FE system of equations.
(As an example, one may consider a triangular element $K$ and choose:
$X(K) = H^2(K)$, $X_h(K)$ -- the space of linear functions over $K$,
and either $H^1$ or $L_2$ norms in \eqref{eqn:eps-a-FEM};
the standard FE approximation error estimates ensue 
\cite{Ciarlet72,Ciarlet80}.)

This serves as a motivation to look for an analogous error 
measure in the FD context. 
It should be emphasized from the outset that this measure is
\textit{local} -- that is, it applies to a single FD grid molecule.

Consider an $N$-dimensional Trefftz space 
$\calT_N(\Omegarm^{(i)})$ ($N \gg 1$) 
corresponding to a given weak-form differential equation over a (small)
domain $\Omegarm^{(i)}$ covering a particular grid molecule.
Equipped with a suitable inner product $[\cdot \, , \cdot]$
and the respective norm,
$\calT_N(\Omegarm^{(i)})$ is a Hilbert space.
A natural choice of such inner product would be the standard one
of $L_2$ or $H^1$, but scaled by a factor $\mathcal{O}(h^{-d})$ in $d = 1,2,3$ 
dimensions, to compensate for the smallness of the volume of $\Omegarm^{(i)}$.

A given FD stencil (i.e. a set of coefficients -- 
for definiteness, real-valued)  
$\underline{s}^{(i)} \in \mathbb{R}^n$ can be viewed 
as a linear functional $g: ~ \calT_N(\Omegarm^{(i)}) \rightarrow \mathbb{R}$.
We focus on the case where $\Omegarm^{(i)}$ is source-free,
so the right hand side of the underlying differential equation and
the FD scheme is zero.

A key idea is to introduce the ``Trefftz'' subspace 
$T_h(\Omegarm^{(i)}) \subset \calT_N(\Omegarm^{(i)})$ of functions 
\textit{satisfying the FD scheme exactly}; that is,
\begin{equation}\label{eqn:Th-defined-via-functional}
\calT_h(\Omegarm^{(i)}): ~~ \{u \in  \calT_N(\Omegarm^{(i)}),
~~  \langle g, u \rangle \,=\, 0 \}
\end{equation}
We shall refer to $\calT_h$ by the acronym SETS: 
\textit{scheme-exact Trefftz subspace}.
By the Riesz representation theorem, the linear functional $g$
can be written as
\begin{equation}\label{eqn:linear-functional-eq-xi-u}
\langle g , \, u(\bfr)\rangle \,=\, 
[g_\bot(\bfr), \, u(\bfr)], \quad 
\forall u \in \calT_N(\Omegarm^{(i)})
\end{equation}
for some $g_\bot(\bfr) \in \calT_N(\Omegarm^{(i)})$. Dependence on the position 
vector $\bfr$ 
is explicitly shown to distinguish functions in $\calT_N$ 
from vectors or functionals.
%
With this notation at hand, the SETS $\calT_h$  can be written as
\begin{equation}\label{eqn:Th-defined-via-inner-product}
\calT_h(\Omegarm^{(i)}): ~~ \{u(\bfr) \in  \calT_N(\Omegarm^{(i)}),
~~  [g_\bot(\bfr), u(\bfr)] \,=\, 0 \}
\end{equation}
For any given $u(\bfr) \in  \calT_N(\Omegarm^{(i)})$, one can then define the
approximation error
\begin{equation}\label{eqn:eps-a-min-T-vs-Th}
\epsilon_a \,=\, \inf_{u_h \in \calT_h(\Omegarm^{(i)})}
\| u - u_h \|_{\calT_h(\Omegarm^{(i)})}
\end{equation}
A critical observation is that the subspace $\calT_h$
does \textit{not} depend on the scaling of the FD scheme, and hence 
the approximation error \eqref{eqn:eps-a-min-T-vs-Th} does not either,
in contrast with the standard consistency error $\epsilon_c$.
The approximation error $\epsilon_a$ is determined 
by solving an orthogonal projection problem; one obtains
\begin{equation}\label{eqn:norm-eps-a-eq-Psi-u-g-Psi-g-g-sq}
\epsilon_a \,=\, \frac{\left| \, [u(\bfr), \, g_\bot(\bfr)] \, \right|}
{\| g_\bot(\bfr) \|_{\calT_N}}
\end{equation}
An explicit expression for $g_\bot(\bfr)$ can be found if a basis in
$\calT_N$ is introduced (see Appendix).

\begin{example}
	For illustration, let us consider one relatively simple example:
	the 2D Laplace equation $\nabla^2 u = 0$ and the five- and nine-point
	stencils shown in \figref{fig:5pt-9pt-stencils-Laplace-2D}.
	(The former is standard, and the latter can be viewed as either
	a Mehrstellen or FLAME scheme \cite{Tsukerman-book07}.)
	
	\begin{figure}
		\centering
		\includegraphics[height=1.2in]{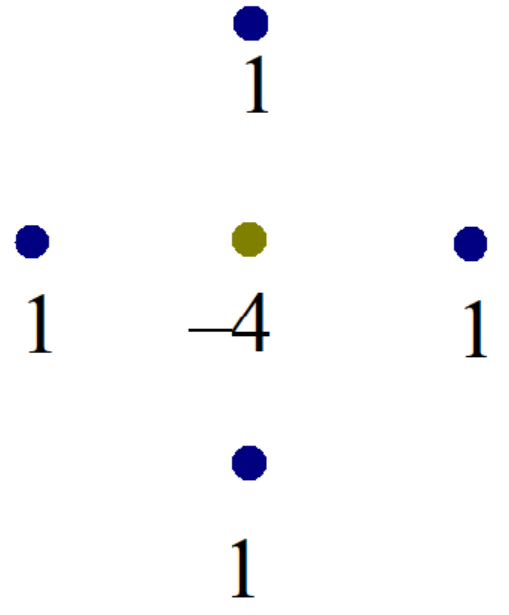}
		\hskip 0.6in
		\includegraphics[height=1.2in]{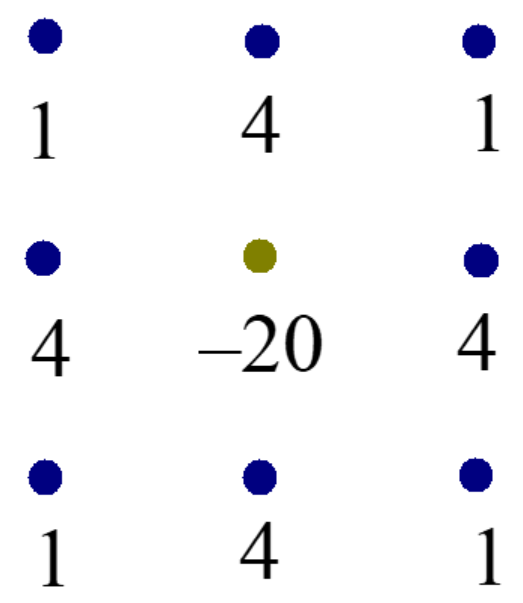}
		\caption{Standard five-point and nine-point stencils
			for the 2D Laplace equation.}
		\label{fig:5pt-9pt-stencils-Laplace-2D}
	\end{figure}
	
The above procedure for evaluating the approximation error
was implemented in symbolic algebra for the following setup:
\begin{itemize}
		\item 
		The Trefftz basis set of $N = 2m+1$ harmonic polynomials
		$\underline{\psi} = \{ 1, \mathrm{Re} \, z^m$, 
		$\mathrm{Im} \, z^m \}$, $m = 1,2, \ldots, M$, $z = x + iy$.
		\item
		The Trefftz space $\calT_N = \mathrm{span} \, \underline{\psi}$,
		equipped with the inner product  $(2h)^{-2} (\cdot \, , \, 
		\cdot)_{L_2}$
		and the respective norm. The factor $(2h)^{-2}$
		compensates for the smallness of the area of the subdomain
		$\Omegarm^{(i)} = [-h, h] \times [-h, h]$.
\end{itemize}
Typical results for the basis set of harmonic polynomials
of orders up to $M = 10$ are summarized in 
\tableref{table:approx-errors-5pt-9pt-Laplace-2D}.

\renewcommand{\arraystretch}{1.5}
\begin{table}
		\caption{The lead terms in Trefftz-space approximation errors
			for the 5-point and 9-point stencils; the 2D Laplace 
	equation.}
		\begin{center}
			\begin{tabular}{c|c|c|c}
				& $h^{-2} \|\cdot\|_{L_2}$ ~&~ $h^{-2} \|\cdot\|_{H^1}$ 
	~&~ $h^{-2} \|\cdot\|_{H^2}$\\
				\hline
				5-point stencil ~&~ $\sim 0.655 h^4$ ~&~ $\sim 4.17 h^3$ 
	~&~ $\sim 
	12.6 
				h^2$ \\
				9-point stencil ~&~ $\sim 1.99 h^8$ ~&~ $\sim 18.5 h^7$ 
	~&~ $\sim 
	98.5 
				h^6$\\
			\end{tabular}
		\end{center}	
		\label{table:approx-errors-5pt-9pt-Laplace-2D}
\end{table}

	
Explicitly, for the standard 5-point scheme 
	(\figref{fig:5pt-9pt-stencils-Laplace-2D}, left) and $M = 4$, 
	one obtains
$$
	g_\bot(\bfr) \,=\, 
	-\frac{1575(x^4 - 6x^2 y^2 + y^4)} {208h^4} 
	- \frac{105}{52}; 
$$
$$
	\| g_\bot(\bfr) \| \,=\, \frac{15 \sqrt{91}}{26}
$$
For the following test function as an example
$$
	f_\mathrm{test} = \exp 2x \cos 2y
	\quad \quad (\nabla^2 f_\mathrm{test} = 0)
$$
the approximation error \eqref{eqn:norm-eps-a-eq-Psi-u-g-Psi-g-g-sq}
is calculated to be
$$
	\epsilon_a \,=\, \frac{16 \sqrt{91}}{315} \, h^4
	\,+\, \mathrm{higher~order~terms}
$$
Repeating this symbolic algebra analysis for
the 9-point scheme (\figref{fig:5pt-9pt-stencils-Laplace-2D}, right)
and $M = 8$, we get 
$\| g_\bot(\bfr) \|_{L_2} = 10725 \cdot 
	665270333^{\frac12}/3440312 \approx 
	80.4$.
(The actual expression for $g_\bot(\bfr)$ itself is too cumbersome 
	to be worth reproducing here.)
For the same test function $f_\mathrm{test}$ as before,
$$
	\epsilon_a \,=\, \frac{256 \sqrt{665270333}}{1045269225} \, h^8
	\,+\, \mathrm{higher~order~terms}
$$
	In all cases the approximation error
	\eqref{eqn:norm-eps-a-eq-Psi-u-g-Psi-g-g-sq}
	is independent of the scaling of the scheme;
	to verify that experimentally, the schemes of
	\figref{fig:5pt-9pt-stencils-Laplace-2D} 
	were multiplied with arbitrary factors,
	and the symbolic algebra results for the error remained unchanged.
	Naturally, though, the approximation accuracy
	does depend on the choice of the norm in the Trefftz space $\calT_N$.
\end{example}

\section{Conclusion}
%
This paper addresses several accuracy-related issues which,
despite their principal theoretical and practical importance 
for FD schemes in electromagnetic applications,
have been overlooked and/or remain controversial.
An important example is the order of Yee-like schemes in the vicinity
of slanted or curved material interfaces: in the existing literature
conflicting claims have been made as to whether this order 
can be higher than one.
This inconsistency is due to several factors: (i) difference in the
accuracy of local and global quantities for FD solutions;
(ii) a very large variety of techniques proposed for accuracy improvement,
and complexity of their mathematical analysis; (iii) the dependence
of the consistency error on the scaling of the scheme
(\sectref{sec:Factors-affecting-analysis}).

The paper focuses on \textit{local} consistency errors and puts forward
two practical methods for an ``apples vs. apples'' comparison of
various difference schemes with arbitrary degrees of freedom,
in homogeneous regions and especially in the presence of material 
interfaces. The first method makes use of \textit{Trefftz test matrices} 
(\sectref{sec:Trefftz-Test-Matrix}), while the second one
relies on a new measure of approximation accuracy in \textit{scheme-exact 
Trefftz subspaces} introduced in \sectref{sec:Approximation-Trefftz-subspaces}.
One particular conclusion is that a loss of local approximation accuracy 
by one order is unavoidable for Yee-like schemes at slanted material
boundaries.
%
\section*{Appendix: An Expression for $g_{\bot}$}\label{app:g-bot}
%
Let $\underline{\psi}(\bfr) = \{ \psi_{\alpha}(\bfr) \}$ 
be a basis in $\calT_N$, arranged as a column vector. 
Then any function $u(\bfr) \in \calT_N$ can be expanded as
\begin{equation}\label{eqn:u-eq-cT-psi}
u(\bfr) \,=\, \underline{u}^T \underline{\psi}(\bfr),
\quad \underline{u} \in \mathbb{R}^N
\end{equation}
where the column vectors are underlined. Euclidean inner product
and that of $\calT_N$ are related as
\begin{equation}\label{eqn:inner-prods-T-RN}
[u(\bfr), v(\bfr)] \,=\, 
[\underline{u}^T \underline{\psi}(\bfr) , \,
\underline{v}^T \underline{\psi}(\bfr)]
\, = \, (\Psirm \underline{u} \, , \, \underline{v})
\end{equation}
Here $\Psirm$ is the Gram matrix of the basis functions:
\begin{equation}\label{eqn:Gram-matrix-psi-psiT}
\Psirm \, = \, [\underline{\psi}, \, \underline{\psi}^T]
\quad \Longleftrightarrow \quad
\Psirm_{\alpha \beta} = [\psi_{\alpha} \, , \, \psi_{\beta}],
~~ 1 \leq \alpha, \beta \leq N
\end{equation}
Applying the $g$ functional -- that is, the difference scheme -- to
each of the basis functions, one obtains the column vector
\begin{equation}\label{eqn:g-psi-eq-g-psi}
\underline{g}_{\psi} \,=\, \langle g \, , \, \underline{\psi}(\bfr) 
\rangle
\, \in \, \mathbb{R}^N
\end{equation}
We can now determine the vector $g_\bot(\bfr)$ representing in $\calT_N$
the functional $g$ and hence the difference scheme itself. We have
\begin{equation}\label{eqn:g-psi-eq-Psi-g-u}
\langle g \, , \, u(\bfr) \rangle
\, \overset{\eqref{eqn:linear-functional-eq-xi-u}}{=} \, 
[g_\bot(\bfr), u(\bfr)] \, \overset{\eqref{eqn:inner-prods-T-RN}}{=}
\, (\Psirm \underline{g} , \, \underline{u})
\end{equation}
Setting in this equation $u(\bfr) = \psi_{\alpha}(\bfr)$,
the respective $\underline{u}$ being the unit vector $e_{\alpha}$,
we derive
\begin{equation}\label{eqn:g-psi-alpha}
\underline{g}_{\psi \alpha} \equiv  \langle g , \, \psi_{\alpha} (\bfr) 
\rangle
\, = \, 
(\Psirm \underline{g} \, , \, e_{\alpha})
\, = \, (\Psirm \underline{g})_{\alpha}
\end{equation}
This gives an explicit expression for $\underline{g}$ -- that is,
for the representation of $g_\bot(\bfr)$ \eqref 
{eqn:norm-eps-a-eq-Psi-u-g-Psi-g-g-sq} in the $\{\psi\}$ basis: 
\begin{equation}\label{eqn:g-eq-inv-Psi-g-psi}
\underline{g} \, = \, \Psirm^{-1} \underline{g}_{\psi}
\end{equation}
As a reminder, $\underline{g}$ in the left hand side is the column
vector corresponding in the chosen $\underline{\psi}$ basis
to $g_\bot(\bfr)$, while the column vector $\underline{g}_{\psi}$
\eqref{eqn:g-psi-eq-g-psi}
contains the values of the difference scheme evaluated on each basis 
function.

For any function $u(\bfr) \in \calT_N$, its orthogonal projection
onto the SETS $\calT_h$ has the form
\begin{equation}\label{eqn:uh-eq-u-minus-alpha-g}
u_h(\bfr) \,=\, u(\bfr) - \alpha g_{\bot}(\bfr)
\end{equation}
where the coefficient $\alpha$ is found from the orthogonality condition
\begin{equation}\label{eqn:u-minus-alpha-g-orth-g}
[u(\bfr) - \alpha g_{\bot}(\bfr), \, g_{\bot}(\bfr)] \,=\, 0
\end{equation}
so that
\begin{equation}\label{eqn:alpha-g-orth-g}
\alpha \,=\, \frac{[u(\bfr), \, g_{\bot}(\bfr)]}
{[g_{\bot}(\bfr), \, g_{\bot}(\bfr)]}
\,=\, \frac{(\Psi \underline{u} \, , \,\underline{g})}
{(\Psi \underline{g}, \,\underline{g})}
\end{equation}

\bibliographystyle{ieeetran}
\bibliography{FDTD_order}

\end{document}